\documentclass[onecolumn]{article}
\usepackage{amsmath,amssymb}

\input xy
\xyoption{all}
\xyoption{arc}
\usepackage[curve]{xy}
\xyoption{poly}
\usepackage{hyperref}
\begin{document}
\vskip 0.4cm
\title{Thermodynamics of a Dust Universe\\
Energy density, Temperature, Pressure and\\
Entropy for Cosmic Microwave Background}
\author{James G. Gilson\quad  j.g.gilson@qmul.ac.uk\thanks{
 School of Mathematical Sciences,
Queen Mary University of London, Mile End Road, London E1 4NS,
United Kingdom.}}
\date{January 24, 2007}
\maketitle
\begin{abstract}
\vspace{0.4cm}
This paper continues the building of the cosmological theory that was introduced in two earlier papers under the title  {\it A Dust Universe Solution to the Dark Energy Problem\/}. The model introduced in this theory has existence before time zero so that it is not necessary to interpret it as of big-bang origin. The location of the Cosmic Microwave Background, within the theoretical structure gives a closing of the {\it fundamentals\/} of the model in terms of the definitions of Temperature, Entropy and other Thermodynamic aspects. Thus opening up a research tool in cosmology in exact agreement with experiment that can compete with the so-called {\it Standard Big Bang Model\/} as a mathematical-physical description of our universe based {\it rigorously\/} on Einstein's general relativity. It is suggested that the singularity at time zero involves a {\it population inversion\/} in the statistical mechanics sense and so justifies the use of negative temperature for the {\it CMB} at negative times. This also has the satisfactory consequence that the Universe's evolution involves entropy steadily increasing over all time from minus infinity through the singularity to plus infinity. An appendix with its own abstract contains an alternative simple classical physics derivation of this model and an extended discussion about how it can be used in the astrophysical context of galactic motions. A cosmological Schr\"odinger equation of great generality is derived which unites cosmology and the quantum description of cosmological objects.
\vskip 0.5cm 
\centerline{Keywords: Dust Universe, Dark Energy, Friedman Equations,}
\centerline{Entropy, Population Inversion, Negative Temperature}
\vskip 0.2cm
\centerline{PACS Nos.: 98.80.-k, 98.80.Es, 98.80.Jk, 98.80.Qc}
\end{abstract}
\section{Introduction}
\setcounter{equation}{0}
\label{sec-intr}
The work to be described in this paper is an extension and general discussion of the significance and physical interpretations of the papers {\it A Dust Universe Solution to the Dark Energy Problem\/} \cite{45:gil} and {\it Existence of Negative Gravity Material. Identification of Dark Energy\/} \cite{46:gil}. The conclusions arrived at in those papers was that the dark energy {\it substance\/} is physical material with a positive density, as is usual, but with a negative gravity, -G,  characteristic and which is twice as abundant as has usually been considered to be the case.  References to equations in those papers will be prefaced with the letter $A$ and $B$ respectively. The work in $A$, $B$, the discussion here and the extensions here have origins in the studies of Einstein's general relativity in the Friedman equations context to be found in references (\cite{03:rind},\cite{43:nar},\cite{42:gil},\cite{41:gil},\cite{40:gil},\cite{39:gil},\cite{04:gil},\cite{45:gil}) and similarly motivated work in references (\cite{10:kil},\cite{09:bas},\cite{08:kil},\cite{07:edd},\cite{05:gil}) and 
(\cite{19:gil},\cite{28:dir},\cite{32:gil},\cite{33:mcp},\cite{07:edd}).
Other useful sources of information have been references (\cite{3:mis},\cite{44:berr}) with the measurement essentials coming from references (\cite{01:kmo},\cite{02:rie},\cite{18:moh}).  Further references will be mentioned as necessary.  After writing $A$ and $B$, I found that Abbe Georges Lema\^itre \cite{47:lem} had produced much the same model in $1927$ but had presented it with a greatly different emphasis and interpretation while also missing important aspects of the significance of the model that have emerged from the version and structure that I had found.  He seems to have been only aware of the positive time solution and then only in a restricted form in his $1927$ paper. See remarks on page $66$ of {\it McVitties's\/} book \cite{50:mcv}. Abbe Georges Lema\^itre is often referred to as the father of the big-bang but it seems to me that this reputation arose out of his omission of the negative time solution in his considerations of this early model. The explicit solution for $r(t)$, equation (\ref{B77.2}) here, is also not mentioned in the $1927$ paper. Up to date, there are generally few references to the explicit solution, (\ref{B77.2}), and sparse developments of its form. In particular, I have found a substantial application of this formula by the authors {\it Ronald J. Adler et al\/} \cite{48:adl} in a paper called {\it Finite Cosmology and a CMB Cold Spot\/}. Their work has common ground with the present paper. However, they fail to realise the power of the scale factor (\ref{B77.2}) and use it in conjunction with a time patching scheme making use of the related looking formula (\ref{1.0}),
\begin{eqnarray}
a(t) \sim  \sinh^{1/2}(t/(2 R_d))\label{1.0}
\end{eqnarray}
where $R_d$ is the de Sitter radius,
to represent the radiation dominated era just as the formula (\ref{2})
is used in the standard model to represent this era. This patching will be discussed in detail in the next section. I have not found who first wrote down the formula (\ref{B77.2}).
The scale factor (\ref{B77.2}) and its universe although known about for many years, seems to have been almost totally eclipsed and overshadowed by the vast effort put into developing over the last eighty years of what is now called {\it The Standard Model\/}.
\section{Defect in Standard model}
\setcounter{equation}{0}
\label{sec-dism}
The standard cosmological model is generally accepted as giving a good and plausible  description of the evolution of the universe from the initially assumed big bang through to the present time and into predicting the future of the universe. At the same time, it can be used to give a convincing case for how the material structures of which the universe is composed now has evolved from simpler and more fundamental materials that were likely predominant at earlier times. This is seen as a building and complexifying process with time which is assumed to be driven by temperature change from the very high temperature in the vicinity of the big bang the low temperate of the present day. The physical processes that are involved can be identified from general earthbound physical theory and technology that associates the need for specific ranges of temperature with the possibility of known processes occurring. Thus the dependence of the temperature $T(t)$ on time, $t$, is the main key to the structural evolution of the universe.

At this moment in history, we have only one convincing theory of gravity, Einstein's General Relativity and its consequence the {\it Friedman equations\/}, as a description for the process of cosmological evolution with epoch. It is generally assumed that the standard cosmological model is rigorously a solution to the Einstein field equations. I do not wish to be dogmatic but I believe the truth of this assumption is questionable. To examine this issue let us consider the mathematical structures that constitutes the  standard model. Nowadays, the standard model is said to describe four main phases in the birth and evolution of the universe. They are (1) inflation and the big bang, (2) a radiation dominated phase, (3) a matter dominated phase and (4) accelerated expansion into the distant future. Essentially three distinct {\it rigorous\/} solutions to Einstein's field equations as represented by their {\it distinct\/} scale factors, $r_A(t), r_B(t), r_C(t)$ and {\it distinct\/} temperature time relations $T_A(t), T_B(t), T_C(t)$ can be found to accommodate the type of physical process assumed to be taking place within each  epoch interval {\it A,B and C\/}. The three scale factors or radii involved are usually represented as,
\begin{eqnarray}
r_A(t)& \sim &  \exp(Ht).\label{1}
\end{eqnarray}
This first $A$ stage involves Guth's inflation determined by a very large value for Einstein's cosmological constant, $\Lambda_A$ and the big-bang event.
\begin{eqnarray}
r_B(t) & \sim & t^{1/2}.\label{2}
\end{eqnarray}
This second stage $B$ is the radiation dominated epoch, with some appropriate value for Einstein's cosmological constant, $\Lambda_B$.
\begin{eqnarray}
r_C(t) & \sim & t^{2/3}.\label{3}
\end{eqnarray}
This third stage $C$ which includes the present time, $t_0= t^\dagger$, is the mass dominated epoch with accelerating expansion into the future determined by the definite {\it measured\/} value of Einstein's cosmological constant, $\Lambda_C=\Lambda$.
I have emphasised that the three solutions {\it A,B,C\/} above are rigorously each a solution of Einstein's field equations. Further, full versions can be found with detailed coefficients so that the $\sim$ sign can be replaced with the $=$ sign. However, these solution are called the {\it Standard Model\/} when they are patched together in time sequence. Mathematically, this patching process can be carried through for both the scale factors and the time temperature relations to find a single form covering the time range $0\rightarrow +\infty$. It seems to me that this final form is not strictly speaking a rigorous solution to Einstein's field equations and for the following reasons. Einstein's field equations are {\it causal\/} in the sense that paths calculated from theory are determined for all time by initial conditions and this applies to the points on the boundary of the expanding universe given by the radial variable $r(t)$. Thus for rigorous solutions $r(t)$ cannot change its form in time passage without some external intervention {\it outside\/} the guiding influence of general relativity. There is no statistical aspect of relativity in its accepted unmodified form that allow for choice between random selection of options at any stage of an evolution process. Statistics is involved in the thermal aspects of modern cosmology but this process the {\it CMB\/} is effectively locked away in a background subspace and can only influence paths through a {\it constant\/} thermal mass contribution in spite of its density changing with temperature and epoch.  If such a change of motion had occurred in actual history, it could be regarded as due to the actual randomness of natural events or as an act of God but this later explanation is certainly outside relativity. Such a change would be on a par with the big-bang concept itself which is generally admitted not to be understood or likely to be explainable by present theory. From the theoretical point of view of the standard model, such changes have been allowed by modern cosmologists in order to conform to human intuitions and preconceptions of how things should be. They may, of course, be right with their intuitive pictures of event sequences but still this is not science and not necessarily proof that the time patched solutions are rigorous solutions to the field equations. This argument against the standard model is strongly reinforced by the fact that solutions of Einstein's field equation involving the cosmological constant requires that this quantity really remains constant over all the time for which the theory or solution is being used. We have seen that the standard model involves at least two definite changes to its parameter values for $r(t)$ and $T(t)$ and $\Lambda$. There is another important aspect of this issue and that is that the model I am proposing is rigorously a solution in isolation. It does not need patching up or joining to other solution to give a full time solution. Further it can supply all the facilities and options that are offered by the standard model for material synthesis resulting from temperature and pressure conditions that change with epoch and more. Showing that this is the case is what this paper is all about.    

\section{Summary of Mathematical Structure of Model}
\setcounter{equation}{0}
\label{sec-sms}
The main theoretical basis for the work to be discussed here are the two Friedman equations that derive from general relativity with the curvature parameter $k=0$ and a positively valued $\Lambda$. For ease of reference these equations and the main results obtained so far from the dark energy model are listed next,
\begin{eqnarray}
8\pi G \rho (t) r^2/3 & = &  {\dot r}^2 - |\Lambda | r^2 c^2/3 \label{B3}\\
-8\pi GP(t) r/c^2 & = & 2 \ddot r + {\dot r}^2/r -|\Lambda | r c^2.\label{B4}\\
r(t) & = & (R_\Lambda /c)^{2/3} C^{1/3} \sinh^{2/3} ( \pm 3ct/(2R_\Lambda)) \label{B77}\\
b & = &  (R_\Lambda /c)^{2/3} C^{1/3}\label{B77.1}\\
C &=& 8\pi G \rho (t) r^3/3\label{B6}\\
R_{\Lambda} &=& |3/\Lambda|^{1/2} \label{B5}\\
\theta _\pm(t) &=& \pm 3ct/(2R_\Lambda) \label{B77.3}\\
r(t) & = &  b\sinh^{2/3}(\theta_\pm (t) ) \label{B77.2}\\
v(t)&=& \pm (bc/R_\Lambda)\sinh ^{-1/3} (\theta _\pm (t))\cosh ( \theta _\pm(t))\label{B80.1}\\
a(t)&=& b(c/(R_\Lambda) )^2\sinh ^{2/3} (\theta _\pm (t))(3 - \coth ^2 (\theta _\pm(t)))/2 \label{B80.2}\\
H(t)&=& (c/R_\Lambda ) \coth(\pm 3ct/(2R_\Lambda)) \label{B81.1}\\
P(t)&=&(-c^2/(8\pi G))(2\ddot r(t)/r(t) + H^2(t) -3(c/R_\Lambda)^2) \label{B81.12}\\
P(t)&\equiv& 0 \ \forall\  t\label{B81.13}\\
P_\Lambda &=& (-3c^2/(8\pi G))(c/R_\Lambda)^2\label{B811.13}\\
P_G &=& (3c^2/(8\pi G))(c/R_\Lambda)^2\label{K811.13}\\
\rho _\Lambda & =& (3/(8 \pi G))(c/R_\Lambda)^2 \label{B81.2}\\
\rho ^\dagger _\Lambda & =& (3/(4 \pi G))(c/R_\Lambda)^2\ =\  2\rho _\Lambda. \label{B81.3}\\
\rho (t)&=&(3/(8\pi G))(c/R_\Lambda)^2(\sinh^{-2}(3ct/(2R_\Lambda))\label{B1.5}\\
&=&3 M_U/(4\pi r^3(t))= M_U/V_U(t).\label{B1.7}\\
\rho _G(t) &=& (G_+  \rho (t)  + G_- \rho ^\dagger _\Lambda)/G\label{B1.9}\\
G_+&=& +G\label{B1.10}\\
G_-&=& -G,\label{B1.11}\\
\ddot r(t) &=& -4\pi r(t)G \rho_G (t) /3\label{B1.12}\\
\ddot r(t)&=& 4\pi r G(\rho^\dagger_\Lambda -\rho (t))/3\ =\ a(t) \label{B99.1}\\
& = &4\pi r^3 G(\rho^\dagger_\Lambda -\rho (t))/(3 r^2)\label{B99}\\
& = &{M^\dagger}_\Lambda G/r^2 - M_UG/r^2\label{B100}\\
M^\dagger_\Lambda & = & 4\pi r^3 \rho^\dagger_\Lambda/3\label{B101}\\
M_U & = & 4\pi r^3 \rho (t)/3\label{B102}\\
P_G (t) /(c^2\rho _G(t)) &=& 1/(\coth^2(3ct/(2R_\Lambda))-3)\ =\ \omega_G(t)\label{B103}\\\omega _\Lambda &=& P_\Lambda /(c^2 \rho _\Lambda)= -1.\label{B104}
\end{eqnarray}
where $M^\dagger_\Lambda$ is the total dark energy mass within the universe and $M_U$ is the total non-dark energy mass within the universe.
Equation (\ref{B1.12}) or equation (\ref{B99.1}) is exactly the classical Newtonian result for the acceleration at the boundary of a spherical distribution  positive $G$  mass density.   
The function $r(t)$ (\ref{B77.2}) is the radius or scale factor\footnote{After writing $A$, I found mention of this scale factor in Michael Berry's Book, p. 129} for this model(\cite{44:berr}). After writing $A$ and $B$, I found the connection with Lema\^itre's work. 

An interesting and significant result that follows easily from this theory is the identification of the Newtonian gravitational Coulomb like potential that is equivalent to Einstein's general relativity with his positive cosmological constant $\Lambda > 0$ or expressed otherwise, the {\it Newtonian\/} Coulomb potential limit of relativity with Einstein's dark energy.
All that is needed is to replace the Newtonian equivalent density for a point source potential at epoch-time, $t$, with the general relativity gravity weighted density, $\rho _G(t)$, equation (\ref{B1.9}),
$$\rho _G(t) =\rho (t) -\rho _\Lambda ^\dagger .$$
This procedure gives the result,
$$\ddot r = -M G/r^2 + r\Lambda c^2/3$$
in place of the Newtonian result
$$\ddot r = -M G/r^2.$$
The inverse square law is simply modified with the addition of a term linear in $r$ and proportional to Einstein's $\Lambda$.
\section{Thermodynamics of the CMB}
\setcounter{equation}{0}
\label{sec-pmd}

The thermodynamics of dust with its characteristically zero pressure as described here by the quantity $P(t)$ equation (\ref{B81.12}) may seem to present this theory with some conceptual difficulties, if it is to include the cosmical microwave background as part of its structure. However, this turns out not to be the case as will be explained. The model is geometrically and dynamically completely defined, it does have pressure as part of its structure but the thermodynamical significance of this pressure which is identically zero over the whole life span of this model and any relation it may have with a temperature is not obvious. The zero dust pressure does decompose into positive and negative signed components and as usual the dark energy material contributes the negative pressure. The dark energy component is identified firmly with truly negatively characterised gravitating material but takes the form of a positive mass density, $\rho ^\dagger_\Lambda$ which is twice the dark energy density, $\rho _\Lambda$, identified by Einstein. There is no reason to believe that the {\it CMB\/} should be directly associated with the dark energy vacuum constituent and also be negative G characterised and so to add the {\it CMB\/} into the structure the obvious choice is to make it part of the conserved mass of the universe which has been denoted earlier by $M_U$, equation (\ref{B102}). This placement for the {\it CMB\/}  is reinforced by the fact that blackbody radiation is not an absolute constant as is the dark energy density but rather depends on temperature which itself is usually assumed to vary with epoch time. I therefore make the strong assumption,
\begin{eqnarray}
M_U =  M_\Delta + M_\Gamma,\label{E1.00}
\end{eqnarray}
where $M_U$, the total conserved {\it non-dark\/} energy mass of the universe,  $M_\Delta$ is the conserved mass that is neither {\it CMB\/} nor {\it non-dark\/} energy mass and  $M_\Gamma$, is the conserved {\it CMB\/} mass of the universe and all are taken to be absolute constants. In this paper, I shall restrict the discussion to the assumption (\ref{E1.00}).
I refer to this assumption as strong as there are other option possibilities all of which can keep $M_U$ costant over all time because the model depends {\it essentially\/} on the assumption that Rindler's constant $C= 2 M_U G$ is an absolute constant and this is what makes the integration of the Friedman equations yield the model. Clearly, $C$ can be kept constant if $M_U$ varies with time with whatever side effects that may have. The main consequence of the assumption (\ref{E1.00}) is that total non-dark energy density, $M_U$  and both $\Delta$ and $\Gamma$ component densities acquire an epoch time dependence as a result of the equations, 
\begin{eqnarray}
M_U &=& \rho (t)V_U(t),\label{E1.10}\\
M_\Delta &=&   \rho_\Delta (t) V_U(t),\label{E1.20}\\
M_\Gamma &=&  \rho_\Gamma (t) V_U(t),\label{E1.30}
\end{eqnarray}
where the volume of the universe at time $t$ is given by $V_U(t) = 4 \pi r^3(t)/3$. Taking the cosmic microwave background radiation to conform to the usual blackbody radiation description, the {\it mass\/} density function for the {\it CMB\/} will have the form
\begin{eqnarray}
\rho_\Gamma (t) &=& a T^4(t)/c^2,\label{E1.40}\\
a&=& \pi ^2 k^4/(15 \hbar ^3 c^3) ,\label{E1.50}\\
 &=& 4\sigma /c,\label{E1.60}
\end{eqnarray}
where $\sigma$ is the Stephan-Boltzmann constant,
\begin{eqnarray}
\sigma  &=& \pi ^2 k^4/(60 \hbar^3 c^2),\label{E1.70}\\
&=& 5.670400\times 10^{-8}\quad W\ m^{-2}\ K^{-4}.\label{E1.80}
\end{eqnarray}
The total mass associated with the  {\it CMB\/} is given by (\ref{E1.30}). That is 
\begin{eqnarray}
M_\Gamma &=&  \rho_\Gamma (t) V_U(t), \label{E1.90}\\
&=& (a T^4(t)/c^2) 4 \pi r^3(t)/3\label{E1.100}\\
&=& (a T^4(t)/(3 c^2)) 4 \pi  b^3\sinh^{2}(\theta_\pm (t)) \label{E1.110}\\
&=& (a T^4(t)/(3 c^2)) 4 \pi  (R_\Lambda /c)^{2} 2 M_U G\sinh^{2}(\theta_\pm (t)) \label{E1.120}\\
&=& (8\pi a T^4(t)/(3 c^4 )) (R_\Lambda )^{2}  M_U G\sinh^{2}(\theta_\pm (t)) \label{E1.121}
\end{eqnarray}
It follows from (\ref{E1.121}) that the temperature, T(t) as a function of epoch can be expressed as,
\begin{eqnarray}
(8\pi a T^4(t) / (3 c^4 ) )  &=& M_\Gamma /((R_\Lambda )^{2}  M_U G\sinh^{2}(\theta_\pm (t)))\label{E1.130}\\
T^4(t) &=& M_\Gamma (3 c^4 ) /(8\pi a (R_\Lambda )^{2}  M_U G\sinh^{2}(\theta_\pm (t)))\label{E1.140}\\
T(t)&=&\pm (M_\Gamma 3 c^4 /(8\pi a (R_\Lambda )^2  M_U G\sinh^2(\theta_\pm (t))))^{1/4}\label{E1.150}\\
T(t) &=&\pm((M_\Gamma 3c^2/(4 \pi a))(r(t)^3))^{1/4}. \label{E1.160}
\end{eqnarray}
The fourth power equation in $T$ has four solutions two of which are real and two of which are complex and so the latter two can be neglected. The positive solution is obviously important but the negative solution  also turns out to play an important role in this theory. I shall return to this issue.  

The well established fact that at the present time, $t^\dagger$, the temperature of the $CMB$ is
\begin{eqnarray}
T(t^\dagger) = T^\dagger =2.725\  K\label{E1.170}
\end{eqnarray}
can be used to simplify the formula (\ref{E1.160}) because  with (\ref{E1.170})
it implies that 
\begin{eqnarray}
0<T(t^\dagger)&=&+(M_\Gamma 3 c^4 /(8\pi a (R_\Lambda )^2  M_U G\sinh^2(\theta_\pm (t^\dagger))))^{1/4}\label{E1.180}
\end{eqnarray}
or
\begin{eqnarray}
T(t)/T(t^\dagger)&=&\pm (\sinh^2(\theta_\pm (t^\dagger)))^{1/4}/(\sinh^2(\theta_\pm (t)))^{1/4}\label{E1.190}\\
T(t) &=&\pm T^\dagger \left(\left(\frac{\sinh(\theta_\pm (t^\dagger))}{ \sinh(\theta_\pm (t)))}\right)^2\right)^{1/4}\label{E1.200}
\end{eqnarray}
where it is understood that the fourth root is taken after the square root.
The formula relating temperature and time (\ref{E1.200}) is different from those which are used in the standard model. The formula here arises in this model in a very natural way. The other thermodynamics quantities for the {\it CMB\/}, free energy F(T,V) as a function of temperature and volume, entropy $S_\Gamma$, pressure $P_\Gamma$ and energy $E_\Gamma$ also arise naturally in their usual forms and as functions of time,
\begin{eqnarray}
F(T,V) &=& - (a/3) V_U(t)T^4(t)\label{E1.210}\\
S_\Gamma (t) &=& \left (-\frac{\partial F}{\partial T}\right)_V= (4 a/3) V_U(t)T^3(t),\label{E1.220}\\
P_\Gamma (t) &=& \left (-\frac{\partial F}{\partial V}\right)_T= (a/3) T^4(t) ,\label{E1.230}\\
E_\Gamma  &=&  a V_U(t)T^4(t).\label{E1.240}
\end{eqnarray}
A very clear and accurate description of the thermodynamics of blackbody radiation can be found in {\it F. Mandl's\/} book \cite{49:man} page 260.
The temperatures and pressures associated with various epoch times associated with this model and labelled alphabetically  are given in the following list. The times have been selected to be only the  frequently mentioned standard  model special times associated with various important physical processes (\cite{52:ham},\cite{54:riz}).
\begin{eqnarray}
Time, secs/yrs &|& Temp, Kelvin\ \ \ |\ \ \   Pressure,\ Nm^{-2}\nonumber\\
t_A=10^{-43}\ s &\rightarrow & 6.521\times 10^{30}\ K,\ \ \ \ 4.56\times 10^{107}\label{E1.260}\\
t_B= 10^{-32}\ s &\rightarrow & 2.062\times 10^{25}\ K,\ \ \ \ 4.56\times 10^{85}\label{E1.270}\\
t_C= 10^{-6}\ \ s  &\rightarrow & 2.062\times 10^{12}\ K,\ \ \ \ 4.56\times 10^{33}\label{E1.280}\\
t_D= 180\ \ \ \ s   &\rightarrow & 1.537\times 10^8\ \ K,\ \ \ \ 1.41\times 10^{17}\label{E1.290}\\
t_E= 3\times 10^5\  yr &\rightarrow &670.4\ K, \quad\quad\quad\ \ \ \ 5.09501\times 10^{-5}\label{E1.300}\\
t_F =10^9\  yr &\rightarrow &11.60\ K, \quad\quad\quad \ \ \ \ 4.57156\times  10^{-12}\label{E1.310}\\
t_G= t^\dagger\approx 14\times 10^9\ yr&\rightarrow &2.728\ K, \quad\quad\quad \ \ \ \ 1.32978\times 10^{-14} \label{E1.320}\\
t_H= 15\times 10^9\ yr &\rightarrow & 2.552\ K, \quad\quad\quad \ \ \ \ 1.07066\times 10^{-14}\label{E1.330}
\end{eqnarray}
These time temperature values are much the same as time temperature values that can be found in the {\it composite\/} standard model and imply temperature ranges that suit various quantum particle physical structure generation processes. They can be found in this theory  with the single temperature {\it all-time\/} formula, $T(t)$, given by equation (\ref{E1.200}).
\vskip 0.2cm 
\centerline{\bf Approximate list of physical processes associated}
\centerline{\bf with temperature ranges implied by $T(t)$:}
\vskip 0.2cm
Preliminary remark: Processes are associated with times in this list using times often quoted in the standard model \cite{54:riz}. The association of process with temperature in the standard model is of very low-level accuracy so I have made no attempt to get accurate connection of time with process. The list is just to show qualitatively that the processes  described in the standard model can also be described in the model that I am proposing with at least equal detail.
The zero and positive valued times listed above from the standard model, $t_A\rightarrow t_H$, interspersed with theoretical values ($0,t_1,t_c,t^\dagger,t_2$) from the present theory, in numerical time order, are given in the following list. The present theory contains time symmetrical event $t\rightarrow -t$ that are not listed.
\vskip 0.2cm
\centerline{Main Events in History of the Universe}
\begin{description}
\item[$t=0$] The {\it singularity\/} when the radial speed is ambiguously, $\pm \infty$
\item[$t_A$] Planck epoch range of super-fast inflation 
\item[$t_B$] Post inflation when there is a hot soup of electrons and quarks
\item[$t_C$] Rapid cooling when quarks convert into protons and neutrons  
\item[$t_D$] Super hot fog of charged electrons and protons impede passage of photons
\item[$t_E$] Electrons protons and neutrons form atoms and photons have free passage
\item[$t_1$] The time when the radial boundary speed of the universe has descended through finite values from $\infty$ to the value $c$
\item[$t_F$] Hydrogen and helium coalesce under gravity to eventually become galaxies
\item[$t_c$] The time when the universes motion changes from
deceleration to acceleration 
\item[$t_G=t^\dagger$] The present time, $t^\dagger$, conditions as they are now
\item[$t_H$] Conditions $10^9$ years into the future
\item[$t_2$] The time when the radial boundary speed of the universe ascends from below to attain the value $c$ again 
\item[$t_\infty =\infty$] 
\end{description}  
\section{Ratio of Radiation Mass to Delta Mass}
\setcounter{equation}{0}
\label{sec-rrd}

The accelerating universe astronomical observational workers \cite{01:kmo} give measured values of the three $\Omega s$, and $w_\Lambda$ to be
\begin{eqnarray}
\Omega_{M,0} &=&8\pi G\rho_0/(3 H_0^2)=0.25^{+0.07}_{-0.06}\label{9}\\
\Omega_{\Lambda,0} &=& \Lambda c^2/(3 H_0^2)=0.75^{+0.06}_{-0.07}\label{10}\\
\Omega_{k,0} &=& -kc^2/(r_0^2 H_0^2) =0,\ \Rightarrow k= 0,\label{11}\\
\omega_\Lambda &=& P_\Lambda/(c^2 \rho _\Lambda) = -1\pm\approx 0.3.\label{11.1}
\end{eqnarray}
According to the strong assumption (\ref{E1.00}) $M_U$, the total conserved mass of the universe, $M_\Delta$ the conserved $\Delta$ mass of the universe and $M_\Gamma$, the conserved {\it CMB\/} mass of the universe are all absolute constants. Thus the ratio of {\it CMB\/} mass to the $\Delta$ mass, $r_{\Gamma,\Delta}=M_\Gamma/M_\Delta$, will be an absolute constant with epoch change together with the same ratio in terms of the corresponding time dependent densities, $\rho _\Gamma (t)$ and $\rho _\Delta (t)$. The value of this ratio is,
\begin{eqnarray}
 r_{\Gamma,\Delta}= \rho_\Gamma (t)/ \rho_\Delta(t) &\approx & 0.00019151\approx 2\times 10^{-4} \label{11.2}\\
\Omega_M(t)  &=& \Omega_\Delta (t)+ \Omega_\Gamma (t)\label{11.21}\\
r_{\Gamma,\Delta}&=& \Omega_\Gamma (t)/ \Omega_\Delta(t) \approx 2\times 10^{-4} \label{11.22}\\
\Omega_M(t)  &\approx& \Omega _\Delta (t) (1 +2\times 10^{-4})\label{11.3}\\
\Omega _\Delta (t) &\approx& \Omega_M(t) (1 -2\times 10^{-4})\label{11.31}\\
\Omega _\Delta (t)  &\approx& \Omega_M(t)_{-0.00005},\ \ \Omega_{M,0} =\Omega _M(t^\dagger) =0.25\label{11.4}\\
\Omega_\Gamma (t) &\approx& \Omega_M(t) \times 2\times 10^{-4},\label{11.41}
\end{eqnarray}
where (\ref{11.21}) is the $\Omega$ equivalent of the strong assumption, (\ref{E1.00}). Hence  (\ref{11.22}) through to (\ref{11.41}).
If we compare (\ref{11.4}) with (\ref{9}), it is clear that the mass weight, $M_\Gamma$, of {\it CMB\/} contribution, $\Omega_\Gamma (t)$,  to the conserved mass contribution $M_U$ is way inside the error limits for $\Omega _{M,0}$. That is to say that as far as mass is concerned, the {\it CMB\/} could be neglected in relation to the $\Omega _\Delta$ contribution as it makes no significant {\it numerical\/} contribution to the {\it basic\/} structure of the theory or the integration process used to derive it. However, it does introduce detail {\it internally\/} to the theory and it is clearly vital to the quantum synthesis of structures story attached to the theory. It reduces homogeneity in an important and useful way  and, as will be shown, it induces a partitioning of the pressure $P(t)$ in the forms for the {\it CMB\/} pressure $P_\Gamma$ and the pressure $P_\Delta$ from the $\Delta$ mass component. This will be addressed next.

The  {\it CMB\/} pressure from equation (\ref{E1.230}) is
\begin{eqnarray} 
P_\Gamma (t) &=& (a/3) T^4(t) ,\label{11.5}\\
             &=& E_\Gamma /(3 V_U(t)) ,\label{11.6}\\
             &=& M_\Gamma c^2/(3 V_U(t))= \rho _\Gamma (t) c^2/3 .\label{11.7}
\end{eqnarray}
From (\ref{B81.13})and following equations the pressure of the dust universe is given by
\begin{eqnarray} 
P(t) &=& P_G  + P_\Lambda \equiv 0 \ \forall\  t.\label{11.71}
\end{eqnarray}
where both $P_G$ and  $P_\Lambda$ are oppositely signed but numerically equal absolute constants. In the previous section, the {\it CMB\/} mass was incorporating into the theory by partitioning the total mass $M_U$ which has been given a fixed numerical value from experiment into the two parts $M_\Delta$ and $M_\Gamma$ while keeping the same constant value for $ M_U $. This defines the $\Delta$ mass, $M_\Delta$, but otherwise makes no difference to the structure of the theory or its correctness as rigorous solution to Einstein's field equations.  However, as the value of the pressure, $P(t)$, is determined by the value within the theory of $M_U$, the non-dark energy part of the universe mass, the partitioning procedure implies a partitioning of the non-dark energy part of the pressure $P(t)$ which is $P_G$ (\ref{11.71}) corresponding to the partitioning of mass, (\ref{E1.00}).

There is an unfortunate anomaly in Friedman cosmology in the perception of pressure and in the way that it is defined. The pressure term in the Friedman equations is assumed to arise basically from the gravitational attraction of the material within the universe. Thus at the universe boundary massive objects will be attracted to within the universe by gravitation and this is regarded as the source of the positive pressure term, $P$, in the Friedman equations. Thus, effectively, $P$, is identified with the pressure that an inflated balloon in equilibrium would produce in reaction to the air pressure, $-P$, within it at equilibrium. It follows that, for consistency, if there is radiation pressure present within the universe it should be placed as a contribution to the $P(t)$ term,  within the theory, with a negative sign. That is we should write for the induced partitioning of $P(t)$
\begin{eqnarray} 
P(t) &=& P_G + P_\Lambda \equiv 0 \ \forall\  t\label{11.72}\\
P_G &=& P_\Delta (t)  - P_\Gamma (t) \label{11.73}
\end{eqnarray}
as the dark energy pressure, $P_\Lambda$, is not involved being not part of an effect from the conserved mass of the universe. The pressure, $P_G$, associated with the gravitational attraction of all the mass within the universe at any time $t$, as has been shown earlier, is an absolute constant while $P_\Gamma (t)$ depends on time. Thus the {\it non-CMB\/} part of the pressure, $P_\Delta (t)$  must also depend on time. Equation, (\ref{11.73}) can be written as
\begin{eqnarray} 
1&=& P_\Delta (t)/P_G  - P_\Gamma (t)/ P_G \label{11.74}\\
r_{P\Gamma G}(t)&=& P_\Gamma(t) /P_G \label{11.75}\\
r_{P\Delta G}(t)&=& r_{P\Gamma G}(t) +1.\label{11.76}
\end{eqnarray}
The ratios $r_{P\Gamma G}(t)$ and $ r_{P\Delta G}(t)$ above give the weight of {\it CMB\/} pressure and the weight of {\it non-CMB\/} to the total non-dark pressure respectively 
   
Let us now consider the relation between the {\it CMB\/} pressure, $P_\Gamma (t)$, and the $\Delta$ mass density,
\begin{eqnarray} 
P_\Gamma (t) &=& (a/3) T^4(t) ,\label{11.8}\\
             &=& E_\Gamma /(3 V_U(t)) ,\label{11.9}\\
             &=& M_\Gamma c^2/(3 V_U(t))= \rho _\Gamma (t) c^2/3  \label{11.10}\\
             &\approx & \rho _\Delta (t) c^2\times 2\times 10^{-4}/3. \label{11.12}
\end{eqnarray}
Equations (\ref{11.72}) and (\ref{11.73}) together, expressed as
\begin{eqnarray} 
P_\Gamma (t) \equiv P_\Delta (t) + P_\Lambda\  \ \forall\  t, \label{11.121}
\end{eqnarray}
gives the important {\it interpretational\/} result that the {\it CMB\/} is in mechanical equilibrium with the $\Delta$ and $\Lambda$ masses combined. This is an alternative characterisation of the dust universe property.
\section{Entropy of the CMB}
\setcounter{equation}{0}
\label{sec-ecmb}
We found at equation (\ref{E1.160}) that there is a choice available for defining physical temperature as a positive or a negative quantity. As we all know, the usual choice in terrestrial physics is the positive one. This choice is made because $S$ is a monotonic increasing function of $E$,
\begin{eqnarray}
\frac{1}{T} = \frac{\partial S}{\partial E}>0.\label{11.873}
\end{eqnarray}
However, there are also physical theory situations where negative values of temperature are encountered and accommodated within conventional thermal physics.  Negative temperatures occur for example in nuclear spin systems. Original work on this aspect is associated with the researchers {\it Purcell and Pound\/} (\cite{53:pap},\cite{49:man}). The second of the previous two references gives much detail about such systems and explanations of their thermodynamics The condition under which negative temperature occurs is often considered to be due to the existence of a sub-system with finite maximum energy that is in weak interaction with the rest of the system so that its own temperature can be defined independently from the rest. The sub-system can then reach equilibrium without being in equilibrium with the whole system. In statistical mechanics, the process involved is sometimes referred to as {\it population inversion\/} with the temperature passing from positive infinity to negative infinity  or conversely at the inversion time. This subsystem property is largely the situation in the cosmological model being presented here where the sub-system is the {\it CMB\/} which has a constant fixed total energy $E_\Gamma$ and so its interaction with the whole universe system which changes markedly with time $t$ must be judged as weak. The property is exactly fulfilled in the present model if, at the singularity time, population inversion actually takes place with a finite negative temperature just before time $t=0$ and reaching $-\infty$ at the singularity then jumping to $+\infty$ immediately after the singularity to then descend through finite positive values to zero as epoch time advances. Such a singularity event depends theoretically on the temperature for negative time being chosen also to be negative and as we have seen we have this option from the theory (\ref{E1.150}).  The infinite temperature jump at $t=0$ can, as is well known, be avoided if one works with the parameter $1/T$ instead of $T$. This is equivalent to regarding inverse temperature as more physically significant than temperature itself.  My feeling is that the idea of a  singularity  population inversion taking place at $t=0$ is a likely scenario. However, I have to admit that the choice of negative temperature for $t<0$ is speculative in spite of it apparent very good fit with the mathematics of the present model. In particular, this assumption does lead to the important conclusion that, if true, the entropy associated with the {\it CMB\/} steadily increases from $t=-\infty \rightarrow +\infty$ and, of course, that is what we would like to be the case. Firm decisions about what happens at the singularity at this time in theory development can only be speculative. Quantum Cosmology when it arrives may hopefully change the situation.       

Starting at equation (\ref{11.870}) is a list of the values of the ratio $ r_{P\Gamma G}(t)$ of {\it CMB\/} pressure to the non-dark energy component  component $P_G$ of the pressure $P(t)$ that arises from the $\Delta$ mass, $ r_{P\Gamma\Delta}(t)$, for fourteen important values of epoch time, $t$.
This model is time symmetric about the time $t=0$, the time of the singularity.
Thus this model has existence before the time $t=0$. This is evident from the form of the radius or scale factor equation, (\ref{B77.2}) provided it is understood that the square is taken before the cube root is taken in the index of the $sinh$ function.  
\begin{eqnarray} 
r_{P\Gamma G }(\pm 0) & \approx &\infty \label{11.870}\\
r_{P\Gamma G }(\pm t_1/10) & \approx & 6.94625 \label{11.872}\\
r_{P\Gamma G }(\pm t_1)&\approx & 0.06493 \label{11.88}\\
r_{P\Gamma G }(\pm t_c)&\approx & 0.00012 \label{11.89}\\
r_{P\Gamma G }(\pm t^\dagger)&\approx & 0.00002 \label{11.90}\\
r_{P\Gamma G }(\pm t_2)&\approx & 2.03 \times 10^{-6} \label{11.91}\\
r_{P\Gamma G }(\pm \infty)& = & 0  \label{11.92}
\end{eqnarray}
The ratio $r_{P\Gamma G}(t)$ is a measure of how the {\it CMB\/} mass pressure relates to the  $P_G$ mass pressure. Towards the singularity, $t=0$, it approaches $\infty$. The $\Delta$ mass pressure ratio to $P_G$ at the times above is given by adding $1$ to the corresponding values above. Thus towards $t=0$ it also approaches $\infty$. Thus from the pressure point of view at the singularity there is {\it mechanical\/} equilibrium between {\it CNB\/} mass  and the $\Delta$ mass and $\Lambda$ mass combined with the contribution from the $\Lambda$ mass being negligible (\ref{11.121}). At the singularity temperature and pressure are both infinite.  This is the equivalent of the idea from big-bang theory that near the singularity the universe is {\it radiation dominated\/}.  
\section{Conclusions}
The cosmological model developed in $A$ and $B$ and further amplified here to show how the thermodynamics of the {\it CMB\/} is included in its structure is {\it rigorously\/} a solution to the full set of Einstein's field equations of general relativity via the Friedman equations.  The model also satisfies {\it exactly\/} the recent measurements by the astronomical dark energy workers. The model need not necessarily be considered to be of the {\it big-bang\/} type because it has a history extending from $t=-\infty$ to $t=+\infty$. However, if the reader cannot accept or just disagrees with the existence of the solution before $t=0$ being joined to the solution after $t=0$, he or she can simply disregard the negative time phase and stick with the big-bang idea. The negative time phase can then be dismissed as an unphysical solution to Einstein's field equations. I shall now give a brief account of the time evolution of the model over the full time range  $t=-\infty$ to $t=+\infty$ and mention a few of the new puzzles thrown up by it. The reader would find it helpful to look at the graphs for velocity and acceleration over this range as I {\it outline\/} the history of events as the model evolves with epoch. See the file darkenergy.pdf \cite{55:gil}, presentation to the {\it 2006 PIRT Conference\/} and the list {\it Main Events in History of the Universe\/} in section {\bf 4}.

The theory implies the following sequence of steps starting at $t=-\infty$. A definite quantity of positively gravitational mass, $M_u$, at density zero uniformly distributed over the whole of an infinitely extended spherical three dimensional Euclidean region of hyperspace is collapsing at a very high speed, $v \gg c$, towards a definite centre. This is the initial situation. Apart from the fact that this mentions the kinematics of a spherical boundary moving towards its centre and as the density of the contents of this collapsing sphere is essentially zero almost nothing {\it physical\/} is happening. Clearly this initial situation is somewhat like the big-bang. However, I think it can be given a plausible explanation with a slight extension of the theory which does not damage its correctness under relativity. I shall come back to this point at the end of this section.   The hyperspace is itself filled with a uniform constant unchanging density of negatively characterised gravitating mass, {\it dark energy\/} from Einstein's positive $\Lambda$. Thus the repulsive gravitational effect of this material within the descending sphere will cause the incoming high velocity of the boundary to steadily be reduced until it reaches from above the value $c$ at the negative time, $-t_2$. All this first phase does present some physical mystery because the boundary motion is superluminal. However, if this early stage motion can be seen as just kinematic that might reduce any detraction by this aspect. Similar problems occur in the standard model. We note that at and above the positive time, $t_2$, we have no information about what physical processes are taking place except that the mass density of the universe is very low indeed which is also obviously the case below and up to the negative time $-t_2$. The next main stage above $-t_2$ and below $-t_1$ is acceptable as within known physics and probably involves physical processes like those occurring above  $t_1$ and below $t_2$. However, it is not necessarily true that the processes in this range are the time-reversed processes in the time symmetric range, displayed in the {\it Events in History\/} list.  Lastly, in the negative time range from $-t_1$ to the singularity at $0$ we are into a mystery superluminal range again but this will likely contain events like those in the positive time range $0$ to $t_1$ but with the same caveat as for the previous range. The sequence of events for the positive time range can be read off from the {\it Main Events in History\/} list which takes us to the end of history at $t_\infty$ with a very low density universe sphere with its boundary rushing to infinity with $v\gg c$.

Returning to the point raised earlier about {\it understanding\/} the initial state, we see that the final state at $+\infty$ is the same as the initial state at $-\infty$ with reversed radial speed of the boundary. This can be given a reasonable explanation by making a reinterpretation of the 3-dimension hyper-space into which the universe is contracting and then expanding after the singularity.  Suppose that the original 3D-hyperspace is replaced with a fixed {\it 3D-surface\/} of a 4D-sphere of very large radius, a well know geometrical trick. Thus the fixed hyper-space becomes a closed 3D-space.  The expanding Universe  can then be regarded as expanding from a point to cover the 4D-sphere surface to some maximum hyper-area and then to contract back over the adjacent hyper-surface to become a point sphere again. This is most easily picture by considering the lower dimensional situation, a circle on an ordinary 3D-sphere surface expanding from a point on the surface to approaching a great circle at great {\it positive\/} speed away from its start point to when, after passing the great circle, descends initially at great negative speed to become an antipodal point.  This is just like the kinematic process in the model is seen to be happening through $t =\pm \infty$. This hyper-spherical interpretation is not part of the model as it stands but it could be incorporated in the structure without damage to the rigorousness of the model as a solution of the Einstein field equations. We can in fact just use this idea to interpret what is happening at infinity and then let the radius of the 4D-hypersphere go to infinity. Its surface then becomes indistinguishable from the original hyperspace.

The main conclusions are that this model for the evolution of the universe has all the advantages that are found in the cosmological standard model together with new perspectives on the nature of dark energy and the amount of it that is present in the background together with a new and clearer representations for temperature, pressure and entropy. Importantly, it does not suffer from the serious defect of the standard model of {\it not\/} being {\it definitely\/} a solution of Einstein's field equations. 
\newpage
\centerline{\Large {\bf Appendix}}
\vskip0.5cm
\large
\centerline{\Large {\bf A Thermal Basis for Gravitational}}
\centerline{\Large {\bf Quantum Vacuum Polarization}}
\centerline{\Large {\bf Descriptive Account of a Fundamental Process}}
\centerline{\Large {\bf in a Friedman Dust Universe}}
\centerline{\Large {\bf with Einstein's Lambda}}
\vskip0.2cm 
\centerline{February 20th 2010}
\vskip 0.75cm  
\section{Appendix Abstract}
A descriptive account is given for the dust universe Friedman lambda model of the universe developed earlier from general relativity by the present author. This description is wrapped around a new and very simple derivation of the model from first principles. The mathematics of this derivation rests on two classical physical equations, the formula for black body radiation and Newton's inverse square law of gravitation, so that without its descriptive wrapping the new derivation which does not involve general relativity directly would occupy about one page of this paper. The descriptive aspect is devoted to showing how the dust universe model can be decomposed into a many subunit form where each galaxy is seen as being a thermal cavity subunit. The time evolution of the whole universe can consequently be seen as being a bundling together of the thermal cavity elements to make up the time evolution structure of the whole universe.  Finally, a cosmological Schr\"odinger equation derived earlier by the present author is significantly generalised to make possible individual quantum state descriptions of the separate galactic thermal cavity elements. Some possible future generalisations of the structure are discussed.       
\vskip0.2cm 
\section{Appendix Introduction}
\setcounter{equation}{0}
\label{sec-aintr}
The work to be described in this paper is an application of the cosmological model introduced in the papers {\it A Dust Universe Solution to the Dark Energy Problem\/} \cite{45:gil}, {\it Existence of Negative Gravity Material. Identification of Dark Energy\/} \cite{46:gil} and {\it Thermodynamics of a Dust Universe\/} \cite{56:gil}. All of this work and its applications has its origin in the studies of Einstein's general relativity in the Friedman equations context to be found in references (\cite{03:rind},\cite{43:nar},\cite{42:gil},\cite{41:gil},\cite{40:gil},\cite{39:gil},\cite{04:gil},\cite{45:gil}) and similarly motivated work in references (\cite{10:kil},\cite{09:bas},\cite{08:kil},\cite{07:edd},\cite{05:gil}) and 
(\cite{19:gil},\cite{28:dir},\cite{32:gil},\cite{33:mcp},\cite{07:edd},\cite{47:lem},\cite{44:berr}). The applications can be found in 
(\cite{45:gil},\cite{46:gil},\cite{56:gil},\cite{60:gil},\cite{58:gil}\cite{64:gil}). Other useful sources of information are (\cite{3:mis},\cite{44:berr},\cite{53:pap},\cite{49:man},\cite{52:ham},\cite{54:riz}) with the measurement essentials coming from references (\cite{01:kmo},\cite{02:rie},\cite{18:moh},\cite{61:free}).  Further references will be mentioned as necessary.
\section{The Dust Universe Model}
\setcounter{equation}{0}
\label{sec-dum}
I have given a detailed mathematical model of the Friedman dust universe in a sequence of papers starting from reference {\it A Dust Universe Solution to the Dark Energy Problem\/}(\cite{45:gil}). This model has structure that involves the current pressing issues arising in modern cosmological theory. The structure of this theory depends heavily on the idea of dark energy and Einstein's Lambda in the context of gravitational vacuum polarization. Every aspect of this theory depends on $\Lambda$ to the extent that, if $\Lambda$ is put equal to zero in the mathematics of the theory, the theory also vanishes. The theory is rigorously a solution to Einstein's field equation and is related to a theoretical structure put forward by Abbe Georges Lema\^itre \cite{47:lem} in $1927$. However, the irony here is that although Lema\^itre is said to be the father of the Big Bang, my version of this theory is not of the big bang type. This oddity suggests that he did not know about the negative time branch of the theory he put forward. My version of this theory is a substantial generalisation of his theory, involving vacuum polarization with Einstein's lambda playing the fundamental role in the structure. The theory has a complete {\it gravitational\/} vacuum theoretical basis which I hope will ultimately merge with the idea of vacuum polarisation in quantum field theory. The mathematics of this vacuum structure as worked out earlier will here be used to set the whole theory onto the idea of a physical {\it emergent\/} from the vacuum picture and at the same time retaining energy conservation intact. A driving motivation has been to construct a theory with roots into the quantum theory of matter and consequently involves the quantum zero-point energy idea. In previous papers this theory has been worked out in great mathematical detail. Here I shall use that mathematical structure as a guide to re-derive the theory from first principles starting from a basic classical thermal equation in the original structure and Newton's inverse square law of gravity. From this new point of view the physics of the system becomes very clear and the basis takes on the form of a more classical intuitively acceptable structure in contrast with its more esoteric and mathematically complex general relativistic form.  
\section{The Thermal Basis}
\setcounter{equation}{0}
\label{sec-ttb}
From classical thermodynamics comes the equation for the {\it mass density\/}, $\rho_\Gamma (t) $, associated with a blackbody radiation system
\begin{eqnarray}
\rho_\Gamma (t) &=& a T^4(t)/c^2, \label{q25q2}\\
a&=& \pi ^2 k^4/(15 \hbar ^3 c^3) ,\label{q25q3}\\
 &=& 4\sigma /c,\label{q25q4}
\end{eqnarray}
where $\sigma$ is the Stephan-Boltzmann constant,
\begin{eqnarray}
\sigma  &=& \pi ^2 k^4/(60 \hbar^3 c^2),\label{q25q5}\\
&=& 5.670400\times 10^{-8}\quad W\ m^{-2}\ K^{-4}. \label{q25q6}
\end{eqnarray}
This well established formula is considered applicable to bounded systems, say a spherical volume of fixed radius, containing electromagnetic radiation. The simplest physical image under which the {\it black body\/} radiation  formula is thought to hold is when the radiation cannot escape but rather is always reflected back towards the spherical interior. The effect is produce a local random field of electromagnetic radiation and consequently to produce a local spatially uniform energy density of radiation at {\it rest\/} within the volume. Assuming Einstein's, $E=m c^2$, law always holds, the consequence is that the mass density above is  generated from the thermal activity within the volume. Thus {\it rest mass\/} is generated from a thermal system despite the fact that the individual photons of which the system is partially composed have no rest mass. Clearly, the mass production process is a consequence of containment. A time varying temperature can occur provided, as indicated, the density function also changes with time. The formula above is said to represent the physics of a thermal cavity so that, if I use the formula above in the astronomical context, it is necessary that I explain what form the appropriate thermal cavity takes in astrophysics.  However, to explain this fundamental facet it is best that I start from what seems to me to be the simplest collection and structure of ideas on which the dust universe model can be based. It has turned out that although in the first place the theory for this model came from Einstein's general relativity it can, alternatively be based on classical Newtonian Gravitational theory with a minor extension. This extension is the addition to the Newtonian concept of gravitating mass which is positive and self attractive the idea that a form of mass can occur which is positive but can be {\it negatively gravitating\/}, in the sense that it is self repulsive and also repels Newtonian mass.  Thus in this theory to distinguish the two types of mass in this work I shall add a plus or minus superscript to specify the type of any positive mass quantity. Thus $m^+$ and $m^-$ are both positive quantities, the first having an attractive gravitation effect on all other masses in its vicinity and the second having a repulsive gravitating effect on all other masses in its vicinity. Clearly the first, $m^+$, is the usual gravitating mass of Newtonian theory. It is also convenient to denote the gravitational acceleration inducing strengths of the two types of particle by $G^+=+G$ and $G^-=-G$ respectively.   It is possible to make  important deductions about the two types without any mathematical theory. The negatively gravitating particles as a consequence of their mutual repulsivity character will spread out,  whereas the positively gravitating particles as a consequence of their mutual attractivity  will tend to clump or form assemblies. Of course both these possibilities will take time for any equilibrium to be reached. The fact that the two types of particle have {\it opposite sign\/} gravitational influence implies that the gravitational interaction between the two types of particle can cause equilibrium configurations between them. This can be explained as follows. Consider a $3$-sphere containing normal gravitating Newtonian particles with a total fixed mass, $M$. The contents of this sphere can contain a variety of mass sized particles distributed within in it in any way. Further let this sphere be immersed in a cloud of negatively gravitating particles, meaning that these negatively gravitating particle can be within or outside the sphere. There will be the two tendencies at work of spreading and clumping with passage of time  conditioned by the spread of the negative gravitating particle being impeded by the gravitational attraction of the positively gravitating particles within the sphere acting against the spread. It seems reasonable to expect an equilibrium state can be reached when the spread is a uniform static field of negative gravitating particles with some arbitrary field of clumped positively gravitating particle within the confines of the sphere which itself could continue to change volume while containing the fixed total of positively gravitating mass, $M$, possibly rearranged. Such rearrangements are possible because only the total mass of the sphere determines its external gravitational field. Assuming the centre of the sphere is its mass centre, an inwards towards the centre of the sphere, source of pressure $P_\Delta$ from the positive gravitating material within the sphere balanced by an outwards pressure $P_\Lambda$ from the negatively gravitating mass material, (\ref{q25q8}), would likely imply an  equilibrium configuration of the type discussed above, 
\begin{eqnarray}
P_\Delta  + P_\Lambda&=&0\label{q25q8}\\
P_\Delta (t)-P_\Gamma (t) +P_\Lambda  &=& 0 \label{q25q9}
\end{eqnarray}
where $P_\Delta $ does {\it not\/} depend on $t$ so that it is consistent with the non time dependence of $P_\Lambda$. However, an equilibrium described by (\ref{q25q8}) does not seem to take into account the thermal photonic field involved in the formula (\ref{q25q2}).
When we do take into account an electromagnetic component as part of the positively gravitating material within the sphere which is the situation being developed here, the component $P_\Delta$ needs to be replaced with $P_\Delta (t) -P_\Gamma (t)$, the $\Gamma$ pressure being preceded with a minus sign as in (\ref{q25q9}) with the $\Delta$ pressure term now also taken to be time dependent. This substitution is to add the negative outward pressure that the photonic pressure would cause. This is necessary as $P_\Gamma (t)$ is defined  as usual as a positive quantity but to agree with the negativity assigned to $P_\Lambda$ it also has to appear with a minus sign. This may seem like sign gymnastics but it all arises because of the convention used in cosmology that positive pressure is taken to cause the inward acceleration, towards source, of normal Newtonian gravity inducing  particles. Physically this addition simply means that the outward photonic pressure makes the normal inward pressure time dependent and shields it so causing the inward normal pressure to be reduced in strength. Thus the photons might be described as contributing pseudo negative gravity. It should also be recognised that even if the pressures concerned are considered to be only within the sphere their effects are transferred via Newton's gravitational law to cause an acceleration field at all positions outside the conceptual sphere defined earlier. This last effect also occurs in general relativity.
In the dust universe cosmology model the theory involved a three dimensional hyper-space in which all the action is played out with time. This space is no geometrical abstraction but rather can be taken to be our familiar three space in which we look out from earth to see the stars and in which the astronomers use their telescopes to survey distant galaxies. We can take this space also to contain a minutely small density distribution of those negatively characterised particles mentioned earlier, just a few such particles per cubic meter. There is probable no more than one such particle within the human body at any moment of time. This same $3$-space will be used here involved as the back ground for the thermal cavity described earlier. Thus generally mass accumulations such as planets, stars and galaxies can be represented by thermal {\it hot spots\/}, more accurately spherical regions or bounded spherical cavities with thermally active interiors, in a sea of negatively gravitating particles at temperatures greater than the temperate given by formula (\ref{q25q2}), this hotspot idea will be explained in more detail later. Thus the thermal cavity can be used to represent partially isolated sub-mass accumulated parts of the whole universe changing under their own dynamics. Thus the whole universe, if such a concept is meaningful, becomes a maximally large thermal cavity. This hot spot idea can be put onto a formal basis using the start formula (\ref{q25q2}) in a modified form that fits this cosmological application, 
\begin{eqnarray}
\rho (t) &=& \rho^\dagger_\Lambda \left(\frac{T(t)}{T(t_c)}\right)^4  \label{q25q7}\\
\rho (t_c) &=& \rho^\dagger_\Lambda \label{q25q81}\\
t_c&=& \frac{2 R_\Lambda}{3 c}\coth^{-1}(3^{1/2}).\label{q125q81}
\end{eqnarray}
In the formula above, $T(t)$ is the temperature at general time $t$ after the singularity at time zero, The time $t_c$ is when the universe's radial acceleration is zero, $\approx 10^{9} yrs$, and $T(t_c)$ is the constant temperature of the negatively gravitational particles whose spatially uniform and time constant mass density function  is given by $ \rho^\dagger_\Lambda $ and has a value that is twice Einstein's dark energy density, $\rho_\Lambda=\Lambda c^2/(8\pi G)$,
\begin{eqnarray}
\rho^\dagger_\Lambda = \frac{\Lambda c^2}{4\pi G}.\label{q25.0q} 
\end{eqnarray}
From (\ref{q25q7}) we see that if the observation time is $t=t_c$ then the mass distribution density, $\rho(t)$, is just equal to the negatively gravitating particle mass density background, $\rho^\dagger_\Lambda$. It also follows that for temperatures $T(t)< T(t_c)$ the region concerned is at a lower temperature than the negatively gravitating particle mass density background or in other words that the temperature is higher relative to absolute zero Kelvin rather than higher relative to the background density. The background is just few degrees Kelvin above absolute zero. We are now in a position to explain the nature of the thermal cavity that underlies the definition of the positively gravitating particle mass density $\rho (t)$, \ref{q25q7}.  Consider the thermal mass density formula (\ref{q25q81}). There are at least two simple ways that this formula can be interpreted. We can describe the density function as representing a condition within a fixed volume $V$ of a variable amount of mass $M(t)$ such as 
\begin{eqnarray}
\rho (t)\implies \rho _{M(t)} = M(t)/V.\label{q25.1.1q} 
\end{eqnarray}
Such an interpretation of the density function $\rho (t)$ could be useful in some contexts but a thermal cavity bounded by the constant volume $V$ would not be a closed dynamical system in any sense because its mass contents would be not conserved with time but necessarily be moving through its boundary. Thus if we are to have a thermal cavity with a conserved mass content, then we must use a different interpretation for $\rho (t)$.
Let us interpret a hot spot as a spherical region centred at the mean position of a total and constant quantity of mass $M$, say, thermally supported according to the formula (\ref{q25q7}) then the density function on the left $\rho (t)$ can be written as $\rho_M (t)$ and then more specifically we have   
\begin{eqnarray}
\rho (t)\implies \rho _M(t) = M/V_M(t), \label{q25.1q} 
\end{eqnarray}
where $V_M(t)$ is a {\it conceptual\/}  spherical space volume containing all the quantity of mass, $M$, and which will clearly have to change with time if $M$ is taken to be constant with the passage of time. I use the term conceptual here because this volume is defined rather than being given as a measurable quantity.  Thus we now have a fixed amount of mass in a varying volume so that we can take the mass M as constituting what in general relativity is often call the {\it substratum material!\/} or a dust like distribution that spreads out uniformly with the changing volume. Of course, if such a volume is expanding then the mass density will get smaller or conversely on contraction the density gets larger. Thus effectively the outer spherical surface of the changing size sphere  represents a spherical boundary suitable for defining the thermal cavity as being the enclosed volume. We can represent the spherical volume $V_M(t)$ in terms of a {\it conceptual\/} time dependent radius vector $r_M (t)$, such that
\begin{eqnarray}
 V_M(t)= \frac{4\pi}{3}r_M^3(t). \label{q25.2q}
\end{eqnarray}
In fact, the  thermal cavity can have any enclosing boundary  which has a centre of volume coincident with the centre of mass of the sphere and the same volume as the sphere because all the volumes involved contain a uniform mass distribution at any given time.  
Using equation (\ref{q25q7}) and its inverted form at (\ref{q25.6q}) and also using equation (\ref{q25.2q}) we have
\begin{eqnarray}
\frac{M}{V_M(t)}&=& \rho^\dagger_\Lambda \left(\frac{T(t)}{T(t_c)}\right)^4 \label{q25.3q}\\
 M&=& M^\dagger_\Lambda (t)\left(\frac{T(t)}{T(t_c)}\right)^4 \label{q25.4q}\\
M^\dagger_\Lambda (t) &=& V_M(t) \rho^\dagger_\Lambda\label{q25.4q1}\\
\frac{4\pi r_M^3(t)}{3 M} &=&\frac{1}{\rho^\dagger_\Lambda} \left(\frac{T(t_c)}{T(t)}\right)^4 \label{q25.5q}\\
r_M^3(t)&=&\frac{3 M }{4\pi \rho^\dagger_\Lambda} \left(\frac{T(t_c)}{T(t)}\right)^4 \label{q25.6q}\\
\frac{\rho_M(t_c)}{\rho_M(t)}=\frac{V_M(t)}{V_M(t_c)}=\frac{r_M^3(t)}{r_M^3(t_c)}&=& \left(\frac{T(t_c)}{T(t)}\right)^4. \label{q25.7q}
\end{eqnarray}
The situation now is that within the time variable spherical volume $V_M(t)$ there is the constant amount of positively gravitating mass $M$ given by (\ref{q25.4q}). Also within this same volume there is the time variable amount of negatively gravitating  mass $M^\dagger_\Lambda (t)$. Thus the radial Newtonian gravitational accelerating field due to both these influences at or just outside the volume at time $t$ is
\begin{eqnarray}
\ddot r (t) &=& -M^\dagger _\Lambda (t) G^-/r^2(t) - MG^+/r^2(t) \label{C19}\\
&=& 4\pi r^3 (t)\rho^\dagger _\Lambda G/(3 r^2(t)) - C/(2r^2(t)) \label{C20}\\
&=& 4\pi r(t) \rho^\dagger _\Lambda G/3  - C/(2r^2(t)) \label{C21}\\
&=& r(t)c^2\Lambda /3 - C/(2r^2(t)) \label{C22}\\
C&=&2 MG^+=2MG \label{C22.1}
\end{eqnarray}
If we multiply equation (\ref{C21}) through by $\dot r$, we obtain
\begin{eqnarray}
\ddot r (t) \dot r(t)  &=& 4\pi r(t) \dot r(t)  \rho^\dagger _\Lambda G/3  - C\dot r(t) /(2r^2(t))\label{C23}\\
\frac{d}{dt}{\dot r(t) }^2/2&=&\frac{d }{dt} r^2(t) \Lambda  c^2/6 + C \frac{d}{dt} r^{-1}(t) /2 \label{C24}\\
{\dot r(t) }^2&=& (r(t) c)^2\Lambda /3 + C r^{-1}(t)   \label{C25}
\end{eqnarray}
The constant of integration that could occur in integrating (\ref{C24}) can be taken to be zero under the conditions that $\dot r(t)$ is taken to be infinite with $r(t)=0$ at $t=0$. Thus the spherical region expands with high speed from the origin, $r(t)=0$ at time $t=0$. 

The solution to the differential equation (\ref{C25}) was obtained in paper $A$ in the form
\begin{eqnarray}
r(t) &=& b\sinh^{2/3}(\pm 3 ct/(2R_\Lambda))\label{C27}\\
R_\Lambda &=& (3/\Lambda)^{1/2}.\label{C28}\\
b & = &  (R_\Lambda /c)^{2/3} C^{1/3}\label{B77.1x}
\end{eqnarray}
and so using (\ref{C27}) with (\ref{q25.7q}) we get the relations
\begin{eqnarray}
\frac{\rho_M(t_c)}{\rho_M(t)}=\frac{V_M(t)}{V_M(t_c)} &=& \left(\frac{T(t_c)}{T(t)}\right)^4 = \left(\frac{\sinh(\pm 3 ct/(2R_\Lambda))}{\sinh( \pm 3 ct_c/(2R_\Lambda)) }\right)^2 \label{q25.8q}\\
\rho_M (t)&=&\rho_\Lambda\sinh ^{-2}(\pm 3 ct/(2R_\Lambda)).\label{q25.81q}
\end{eqnarray}
At this point the problem of finding the detailed description of the model is completely solved the fundamental equation for this solution, (\ref{C27}), is the formula for the radius of the thermal cavity at time $t$. From this formula all elements of the structure of this theory can be obtained by differentiation or algebraic manipulation as for example (\ref{q25.8q}) and (\ref{q25.81q}) were obtained. In particular, Hubble's time dependent space-wise constant is obtained as 
\begin{eqnarray}
H(t)= \frac{\dot r_M (t)}{ r_M (t)}= (c/R_\Lambda ) \coth(\pm 3ct/(2R_\Lambda)) .\label{q25.82q}
\end{eqnarray}
\section{Cosmological Thermal Cavities}
\setcounter{equation}{0}
\label{sec-ctc}
Above I showed that the dust universe model can alternatively be derived from simple classical theory and  expressed its structure in terms of thermal cavities. The first question that arises is, how can this alternative theory be used in the astronomical context? The thermal cavity construction structure introduced above is a rigorous consequence of general relativity and it is also a local theory rather than a global cosmological theory. Local in the sense that it can be used to replace the usual global cosmology type  by an {\it assembly\/} of {\it thermal cavities\/}. All such thermal cavities can be taken to be expanding as part of a global substratum. Thus there is an obvious choice for one type of physical representation of these thermal cavities. This first choice is the galactic unit. Galaxies on the average, isolated from forces other than gravity, move with the substratum flow. In fact, their motion can be regarded as defining the substratum flow. Clearly, much space wise smaller astronomical objects could be used to define astronomical thermal cavities but I shall confine the discussion here to the galactic unit. The other characteristics of galactic existence are their overall identity persistence over very long periods of epoch-time with constant total mass to a high order of accuracy. 
The $\rho (t)$ that I am using can be explained by studying the three equations from earlier repeated below
\begin{eqnarray}
\rho (t)&=& \rho_\Lambda\sinh ^{-2}(\pm 3 ct/(2R_\Lambda)).\label{q25.91q}\\
\rho (t)\implies \rho _M(t) &=& M/V_M(t) \label{q25.89q}\\
M/V_M(t)&=&\rho_\Lambda\sinh ^{-2}(\pm 3 ct/(2R_\Lambda)).\label{q25.90q}
\end{eqnarray}
The equation (\ref{q25.91q}) for $\rho (t)$ from the dust universe model sums up accurately  all that is known physically about the expansion of the universe with epoch time, $t$. It is thus a very fundamental quantity and, curiously, apart from its dependence on $t$ it only has a dependence on Einstein's $\Lambda$, the velocity of light $c$ and $G$ through Einstein's dark energy density, $\rho_\Lambda$. Importantly, it also does not depend on position, it is spatially homogeneous. Notably, this formula contains no mention of any quantity of mass , $M$, there is no such involvement at all. I first introduced this formula as a description of the density evolution for the whole universe. However, having found that this formula can be derived from Newton's theory of gravity it became clear that it is applicable to the problem of the time evolution of sub-regions of the universe. Clearly a galaxy is a large sub-region of the universe so that galaxies are natural structures on which this formula can be tested. In the second formula above I introduce a density $\rho_M(t)$ and take it equal to $\rho (t)$, the density of the universe and this then implies equation (\ref{q25.90q}) from which, given a definite amount of mass, $M$, the definite time dependent mass associated volume, $V_M(t)$, can be found. Thus if $M=M_U$, the mass of the universe, we get the time evolution density of the universe $\rho_{M,U}(t)$ and if $M=M_{galaxy}$, the mass of a galaxy, we get the time evolution of the galaxy density, $\rho_{M,galaxy}(t)$  from the formula. It is this associate definite time dependent volume associated with a galaxy of known mass that I take to be the volume of its bounding thermal cavity. From this discussion it follows that the mass densities of galaxies are all equal at the same cosmological time $t$ for this model, but they do differ in mass content and volume with the definition of volume that I am using. This leads to a very convenient explanation for how the galaxies became separated off from each other as isolated masses over time.  Up to now, I have taken the mass $M$ contained within this conceptual volume $V_M(t)$ to be uniformly distributed within the boundary of this volume, an assumption involved in the generation of formula (\ref{q25.91q}). However, I have also taken this total contained mass to be positively gravitating and so that {\it physically\/} it will clump with time passage and cease to be uniformly distributed. Thus the expanding volume boundary will separate from the contained mass within, so that with sufficient time this mass will occupy a volume, $V_{M,clumped}(t)$, substantially less than $V_M(t)$. Measurements will likely  pick up the volume, $V_{M,clumped}(t)$, for a galaxy and so it will be ascribed a physical density larger than $\rho_M(t)$ assuming the mass has been measured accurately. Returning to the question of galactic distribution, we can infer that if initially at time just greater than the singularity time, $t=0$, at which time it becomes possible to talk about volume, we can assume all the mass in the universe was uniformly distributed over the then small volume. The volume of the universe  could then be considered partitioned into a very large number $n$ of contiguous randomly different sized sub-volumes, very approximately $n\approx 10^{11}$, say, with the uniform mass distribution remaining unchanged by the partitioning. Every sub volume would have the same density as the whole universe. These sub-volumes with their various valued constant with time mass content could have been the seeds of the galaxies we see today having evolved with time according to the formula, (\ref{q25.90q}), with additionally clumping to their separate centres of mass and with density only changing with the epoch time in unison with the density change of the whole universe with time. There is no provision for clumping in the original formula (\ref{q25.90q}) and there is no need for their to be. This formula simply expresses the same Newtonian gravitational feature that given a spherical region centred at the centre of mass of a spatially {\it variable\/} mass density of total mass $M$, the gravitational field at the boundary or just outside this boundary is the same as would exist if the mass $M$ was {\it uniformly\/} distributed over the volume within the same boundary. It is tempting to refer to a thermal cavity that represents a galaxy as a three space hot spot or hot region but to do this it seems we need to use a physically measured temperature, $T_{phys,galaxy}(t)$, say for which we can say within the region of the galaxy  $T_{phys,galaxy}>T(t)$. This is  because  $T(t)$ like $\rho (t)$ is a uniform over space quantity as can be seen from (\ref{q25.8q}) and also is a $3$-space temperature giving the temperature throughout the universe of all elements of the moving positively gravitating substratum which does not take into account local variations of temperature that would occur due to the presence of a galaxy which will be clumped with time. However, it is necessary to find some way of importing variable space position mass density distributions into the theory in order to include the effects of clumping. A way that this can be achieved is shown in the next section.
\section{Cosmological Schr\"odinger Equation}
\setcounter{equation}{0}
\label{sec-cse}
In reference (\cite{64:gil}), I showed that the whole theory for the dust universe model can be obtained as a quantum density $\rho_{nl}(t)$ from the standard general Schr\"odinger equation (\ref{z3}) with the condition $\nabla^{2} \Psi_{nl,\rho} ({\bf r},t) =0$ and the external potential $ V({\bf r,t})$ replaced with the feed back term $V_C (t)$ given at (\ref{z1}).
\begin{eqnarray}
\frac{i\hbar\partial \Psi_{nl,\rho} (t) }{\partial t} &=& -\frac{\hbar ^2}{2m} \nabla^{2} \Psi_{nl,\rho} ({\bf r},t) +(V_C (t) ) \Psi_{nl,\rho }(t)
\label{z0}\\
V_C (t)&=& -(3i\hbar/2)H (t)\label{z1}\\
H(t)&=& (c/R_\Lambda ) \coth(\pm 3ct/(2R_\Lambda)) \label{z2}\\
i\hbar \frac {\partial \Psi ({\bf r},t)}{\partial t} &=& -\frac{\hbar ^2}{2m} \nabla^{2} \Psi({\bf r},t) +V({\bf r,t}) \Psi({\bf r},t). \label{z3}
\end{eqnarray}
$H(t)$ above is the Hubble function from the dust universe theory. Here I shall generalise equation (\ref{z0}) by instead of replacing $V({\bf r,t})$ with $V_C (t)$ in (\ref{z3}) , I shall add it and drop all the subscripts to give
\begin{eqnarray}
\frac{i\hbar\partial \Psi ({\bf r,t}) }{\partial t} &=& -\frac{\hbar ^2}{2m} \nabla^{2} \Psi ({\bf r},t) + V({\bf r,t}) \Psi ({\bf r},t) +V_C (t) \Psi ({\bf r},t).\label{z4}
\end{eqnarray}
I now claim that the Schr\"odinger equation (\ref{z4}) is a general cosmological Schr\"odinger equation. Clearly it differs from the normal general Schr\"odinger equation at (\ref{z3}) in that it has a special external potential of the form $ V_S({\bf r,t}) =V({\bf r,t}) +V_C (t)$ and is only non general in the very weak sense that it has an additional time only dependent part. Let us now consider solutions to the cosmological Schr\"odinger equation (\ref{z4}) with the special form
\begin{eqnarray}
\Psi ({\bf r},t)= \Psi_1 ({\bf r},t) \Psi_{nl,\rho} (t) ).\label{z5} 
\end{eqnarray}
Substituting this form into (\ref{z4}) and using the equation for $\Psi_{nl,\rho} (t)$ at (\ref{z0}) we get the result
\begin{eqnarray}
\frac{i\hbar\partial \Psi_1 ({\bf r,t}) }{\partial t} &=& -\frac{\hbar ^2}{2m} \nabla^{2} \Psi_1 ({\bf r},t) + V({\bf r,t}) \Psi_1 ({\bf r},t).\label{z6} 
\end{eqnarray}
Thus we have recovered the original general Schr\"odinger equation for the factor $\Psi_1 ({\bf r},t) $ in equation (\ref{z5}). Consequently this factor can be any solution of that equation. Clearly the  cosmological solutions of the cosmological Schr\"odinger equation embraces all  possible quantum theory solutions to the usual general Schr\"odinger equation (\ref{z6}). The cosmological Schr\"odinger equation has the usual probability density that goes along with its solution of the form
\begin{eqnarray}
\rho_S ({\bf r},t)= \Psi ({\bf r},t) \Psi ^* ({\bf r},t)= \Psi_1 ({\bf r},t) \Psi_1 ^* ({\bf r},t) \Psi_{nl,\rho }^2(t)
\label{z7} 
\end{eqnarray}
where  $\rho_S ({\bf r},t)$ is a cosmological mass density with position variability of great generality which can involve all known solutions to the standard Schr\"odinger equation  and any other solutions yet to be found. So that on this basis models of the universe can be found involving individually described galaxies of any known quantum internal structure bundled together to describe the whole universe in great detail. The model will allow galaxies to be described as cosmological hotspots.
 \section{Appendix Conclusions}
\setcounter{equation}{0}
\label{sec-conc}
The theoretical structure described above represents a strong unification of quantum mechanics and cosmology through the amalgamation of the cosmological Schr\"odinger equation from the Friedman equations originating general relativity and the general Schr\"odinger equation with an external potential. The generality of the Schr\"odinger structure is in no way compromised by this unification so that the new cosmological version has all the variety of solutions available for cosmology as does its non cosmological form. What the theory does is to set up a cosmological platform strongly linked to the cosmological substratum which enables the mass content of the platform to be described in quantum mechanics terms. The cosmological Schr\"odinger equation above can be used to describe the whole or part of the universe and it can be involved in some obvious generalisations of the theory. The time evolution theory structure for any unit can be given it own time shifted coordinate so that its singularity moment has a distinctive value and it can also be assumed to originate from a distinct position in hyper-space. Many such elements could be bundled to give a different version of the total universe described above. Statistical assemblies of such bundles could be set up and then open the way for a more general cosmology based on quantum statistical mechanics. Such a move into statistical mechanics would ameliorate the difficult idea of a singularity at time zero by at least the same order of magnitude that success at searching for one needle in a haystack is improved by searching for $10^{11}$ needles in the same haystack.    The new short derivation of the dust universe model is a surprising result as it depends only on classical concepts while confirming a structure from general relativity involving Einstein's $\Lambda$ term. This seems to me to reinforce the validity and correctness of the $\Lambda$ term apparently about which Einstein had such doubts. 
\vskip 0.5cm
\leftline{\bf Acknowledgements}
\vskip 0.5cm
\leftline{I am greatly indebted to Professors Clive Kilmister and} 
\leftline{Wolfgang Rindler for help, encouragement and inspiration}

\end{document}